\documentclass[reprint,superscriptaddress,nofootinbib,amsmath,amssymb,aps]{revtex4-2}
\usepackage[utf8]{inputenc}
\usepackage{amsmath}
\usepackage{graphicx}
\usepackage{dcolumn}
\usepackage{bm}
\usepackage{soul}
\usepackage{color}

\newcommand{\BEEC}{EEC$_{\text{DIS}}$ }
\newcommand{\bn}{\bar{n}}

\newcommand{\mathI}{\mathcal{I}}
\newcommand{\bmat}[1]{\boldsymbol{#1}}

\newcommand{\lb}{\Big{\lbrack}}
\newcommand{\rb}{\Big{\rbrack}}
\newcommand{\lp}{\Big{(}}
\newcommand{\rp}{\Big{)}}

\newcommand{\nl}{\nonumber \\[5pt]}
\newcommand{\eq}[1]{eq.~(\ref{eq:#1})}

\begin{document}
\title{Energy-energy correlators in Deep Inelastic Scattering}

\author{Hai Tao Li}
\email{haitao.li@northwestern.edu}
\affiliation{HEP Division, Argonne National Laboratory, Argonne, Illinois 60439, USA}
\affiliation{Department of Physics \& Astronomy, Northwestern University, Evanston, Illinois 60208, USA}
\affiliation{Los Alamos National Laboratory, Theoretical Division, Los Alamos, NM, 87545, USA}
\author{Yiannis Makris}
\email{yiannis.makris@pv.infn.it}
\affiliation{INFN Sezione di Pavia, via Bassi 6, I-27100 Pavia, Italy}
\author{Ivan Vitev}
\email{ivitev@lanl.gov}
\affiliation{Los Alamos National Laboratory, Theoretical Division, Los Alamos, NM, 87545, USA}

\begin{abstract}
The energy-energy correlator (EEC) is an event shape observable which probes the angular correlations of energy depositions in detectors at high energy collider facilities. It has been  investigated extensively in the context of precision QCD. In this work, we introduce a novel definition of EEC adapted to the Breit frame in deep-inelastic scattering (DIS). In the back-to-back limit, the observable we propose is sensitive to the universal transverse momentum dependent (TMD) parton distribution functions and fragmentation functions, and it can be studied within the traditional TMD factorization formalism. We further show  that the new observable is  insensitive to  experimental pseudorapidity cuts, often imposed in the Laboratory frame due to detector acceptance limitations. In this work the singular distributions for the new observable are obtained in soft collinear effective theory  up to $\mathcal{O}(\alpha_s^3)$ and are verified by the full QCD calculations up to $\mathcal{O}(\alpha_s^2)$. The resummation in the singular limit is performed up to next-to-next-to-next-to-leading logarithmic accuracy. After incorporating non-perturbative effects, we present a comparison of our predictions to  PYTHIA 8 simulations. 
\end{abstract}

\maketitle
\section{Introduction}
\label{sec:intro}

Event shape observables (such as thrust, C-parameter, etc.) are measures of  the energy flow,  multiple particle correlations, and the radiative patterns in high energy collisions. They have been extensively investigated at various colliders and, over the past several decades, have played a central role in our understanding of the perturbative and  non-perturbative dynamics of Quantum Chromodynamics (QCD). 

The energy-energy correlator (EEC) is such an observable, which was originally introduced in the context of $e^+e^-$ collisions as an alternative to the thrust family of event shapes. EEC is defined as follows~\cite{PhysRevLett.41.1585}, 
\begin{align}
    \text{EEC} =  \sum_{a,b} \int \frac{d\sigma_{e e\to a+b+X}}{\sigma}\, w_{ab}\, \delta(\cos\chi_{ab}-\cos\chi) \, , 
\end{align}
where the sum runs over all the hadron pairs $(a,b)$ and the cross section is weighted by $w_{ab} \equiv 2E_a E_b /s$, which is the product of the energies of $a$ and $b$ normalized by the center-of-mass energy of the system. EEC measures the energy correlations as a function of the opening angle $\chi_{ab}$ between particles $a$ and $b$. Significant theoretical effort has been devoted to this observable, including fixed-order
calculations~\cite{Belitsky:2013xxa,Belitsky:2013bja,Belitsky:2013ofa,deFlorian:2004mp,Dixon:2018qgp,Gao:2020vyx,Luo:2019nig,Henn:2019gkr,Tulipant:2017ybb}, and QCD factorization and resummation in the back-to-back ($\chi \to \pi$)  and collinear  ($\chi \to 0$) limits~\cite{Moult:2018jzp,Dixon:2019uzg,Kologlu:2019mfz,Korchemsky:2019nzm,Chen:2020uvt}. In the back-to-back limit EEC is related to the transverse momentum distributions (TMD) and  very recently~\cite{Ebert:2020sfi}, was calculated at N$^3$LL$^\prime$ accuracy after obtaining the $\mathcal{O}(\alpha_s^3)$ singular distributions.

At hadronic colliders, an adaptation of EEC known as transverse-energy-energy correlation (TEEC) considers only the momenta in the transverse plane in order to construct the corresponding observable~\cite{Ali:1984yp}. TEEC was calculated including  next-to-leading order (NLO) QCD corrections in ref.~\cite{Ali:2012rn}, and NNLL resummation in the dijet limit was accomplished in ref.~\cite{Gao:2019ojf}.

For observables like EEC and TEEC  soft radiation contributes only through recoil to the energetic collinear particles, since direct contributions from soft emissions are suppressed by the energy weighting factor.  Thus, we should anticipate to have smaller non-perturbative corrections compared to other event shape variables. Moreover, owing to the high perturbative accuracy achieved both in resummed and fixed order calculations~\cite{Moult:2018jzp,Gao:2019ojf,Li:2020bub,Ebert:2020sfi}, complemented  by high precision measurements~\cite{Akrawy:1990hy,Decamp:1990nf,Adeva:1991vw,Abe:1994wv,ATLAS:2015yaa,Aaboud:2017fml,ATLAS:2020mee}, EEC and TEEC observables offer the opportunity for precision studies in QCD.  In particular, EEC and TEEC have been used for precise extractions of the strong coupling constant. For a recent review see section 9 of ref.~\cite{10.1093/ptep/ptaa104}.  

Despite its important applications to precision QCD and the connection of the EEC observable to TMDs, little has been done in the context of  Deep Inelastic Scattering (DIS). With the future Electron-Ion-Collider (EIC) ~\cite{Aschenauer:2014cki} on the horizon, progress in this direction is urgently needed. The first study of TEEC in  DIS was performed in~\cite{Li:2020bub}, where the observable was defined as the correlation between final state hadrons and the scattered lepton in the transverse plane of the Laboratory frame. Furthermore,  in ref.~\cite{Ali:2020ksn} the original definition of TEEC and its asymmetry were also  considered in a fixed-order study in the Breit frame. 

In this work we introduce a new definition of EEC in the Breit frame, which is the natural frame for the study of Transverse Momentum Dependent (TMD) physics~\cite{Collins:2011zzd}.
In this frame, the target hadron moves along $\hat{z}$ and the  virtual photon  moves in the opposite direction. The Born-level process is described by the lepton-parton scattering $e+q_i\to e+q_f$, 
where the outgoing quark $q_f$ back-scatters in the direction opposite to the proton.  Hadronization of the struck quark will form a collimated spray of radiation close to the $- \hat{z}$ direction. In contrast, initial state radiation and beam remnants are moving in the opposite direction close to the proton's direction of motion.  It is this feature of the Breit frame, which leads to the clean separation of target and current fragmentation that we utilize to construct the novel EEC observable in DIS.

We denote the new event shape variable EEC$_{\text{DIS}}$ to avoid confusion with the conventional observable. Our definition reads,
\begin{equation} \label{eq:definition-DIS}
    \text{EEC}_{\text{DIS}}= \sum_a \int\, \frac{d\sigma_{e p \to e+ a+X}}{\sigma}\, z_a\, \delta(\cos\theta_{ap} - \cos\theta)\;,
\end{equation}
where
\begin{equation}\label{eq:weigth-DIS}
    z_a \equiv \frac{P\cdot p_a}{P\cdot (\sum_i p_i)}\;,
\end{equation}
and $p_a^{\mu}$ and $P^{\mu}$ are the momenta of the hadron $a$ and the incoming proton respectively. The angle $\theta_{ap}$ is the polar angle of hadron $a$, which is measured w.r.t. the incoming proton.\footnote{Note that what we propose is different from the discussion in~\cite{Ali:2020ksn}, where a fixed order QCD calculation for TEEC (as defined for hadronic colliders) was performed in the Breit frame DIS for the dijet configuration. This observable exhibits very different characteristics  and is  suppressed by $\alpha_s$ compared to what we propose here.}   Note that the asymmetric weight function, $z_a$, is Lorentz invariant and is suppressed for soft radiation and radiation close to the beam direction. Furthermore, this definition of \BEEC naturally separates the contribution to the $\cos\theta$ spectrum from: i) wide angle soft radiation, ii) initial state radiation and beam remnants, and iii)  radiation from the hadronization of the struck quark. This unique feature makes the new observable in the back-to-back limit ($\theta \to \pi$) insensitive to experimental cuts on the particle pseudorapidity (in the Laboratory frame) usually imposed due to detector acceptance limitations in the backward and forward regions, making the comparison of theory and experiment in this region even more accurate. This definition of EEC is spherically invariant, however, we discuss in the appendix~\ref{sec:BEEC-LI} a possible definition that is fully Lorentz invariant and can be measured directly in any frame.   

For $\pi-\theta \sim 1 $ the distribution is very well described by the fixed order QCD calculations, while in the back-to-back limit resummation of enhanced logarithms is required for reliable predictions. To this end, in the back-to-back limit, the cross section can be factorized, within the soft-collinear effective theory (SCET) framework~\cite{Bauer:2001yt, Bauer:2001ct, Bauer:2000yr, Bauer:2000ew,Beneke:2002ph}, as a convolution of TMD beam, soft,  and TMD fragmentation functions, which share the same operator definitions as in the conventional single hadron semi-inclusive DIS  factorization. This ensures the universality of TMD  parton distribution functions (TMD PDFs) appearing in \BEEC when we compare to other observables, such as  semi-inclusive DIS and jet-TMDs~\cite{Gutierrez-Reyes:2018qez,Gutierrez-Reyes:2019vbx,Gutierrez-Reyes:2019msa, Arratia:2020ssx} in the Breit frame. Moreover, through an operator product expansion (OPE) the TMD beam or TMD fragmentation function can be matched to the collinear PDF and collinear fragmentation function, respectively. As discussed in ref.~\cite{Moult:2018jzp}, after summing over all the final state hadrons, the collinear fragmentation functions can be removed using momentum sum rules and, therefore, the only hadronic matrix elements that enter the factorization in the back-to-back limit are the TMD PDFs. As a result, \BEEC provides a novel approach to TMD physics, avoiding the dependence on fragmentation functions. An important aspect of the EEC observable is that the hadronization corrections are incorporated only at the transverse momentum level through the EEC jet function. This is in contrast to the conventional SIDIS measurements, where one needs to include also the strong hadronization effects introduced at the energy fraction variable, $z_h$, of the identified hadrons. We expect that this feature of EEC will help moderate the correlation of TMDPDFs with hadronization effects in phenomenological extractions.

In this work we construct the resummed cross section up to N$^3$LL accuracy and merge onto the NLO ($\mathcal{O}(\alpha_s^2)$) fixed order QCD result. To achieve the resummed result,  we use the  four-loop cusp and the three-loop hard, soft, and jet anomalous dimensions. The fixed order singular results for the various elements of the factorization formula we take from the literature. The TMD PDFs have been calculated up to three loops~\cite{Gehrmann:2012ze, Gehrmann:2014yya,Luebbert:2016itl,Echevarria:2016scs,Luo:2019hmp, Luo:2019bmw,Luo:2019szz, Luo:2020epw,Ebert:2020yqt}, and the soft function at the same accuracy can be found in refs.~\cite{Li:2016ctv,Moult:2018jzp}. The jet function is the second Mellin moment of the TMD fragmentation functions which are also available up to N$^3$LO~\cite{Echevarria:2016scs, Echevarria:2015usa, Luo:2019hmp, Luo:2019bmw,Luo:2020epw,Ebert:2020qef}. The fixed order QCD result is obtained numerically using  \textsc{NLOJet}\small{++}~\cite{Nagy:2001xb}. We also perform a consistency check of our approach by comparing the LO and NLO singular distributions, obtained by the factorization formula, to the full QCD prediction in the back-to-back limit.  

The paper is organized as follows. In Section~\ref{sec:definition} we introduce the relevant modes of EEC$_{\rm DIS}$ in SCET and the factorized formula in the back-to-back limits. In addition, we discuss  the hadronization corrections. In Section~\ref{sec:numerics} we present our predictions up to N$^3$LL+NLO and compare our results with PYTHIA simulations. Section~\ref{sec:subsets} shows how for the  EEC observables we can select a subset of hadrons in final state, to be used for flavor tagging in TMD PDFs and TMD FFs. We conclude in Section~\ref{sec:concl}. We propose a Lorentz invariant definition of EEC in DIS in the Appendix.
\begin{figure*}
    \centering
    \centerline{\includegraphics[width = 0.8 \textwidth]{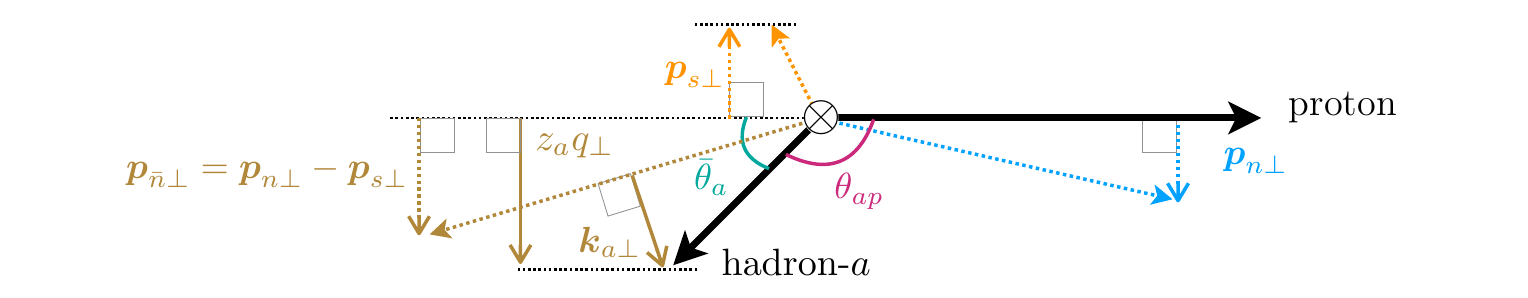}}
    \caption{\label{fig:measurment} Illustration of the measurement of the transverse momentum $\bmat{q}_{\perp}$ of the hadron-$a$ w.r.t. the proton axis in the Breit frame.}
\end{figure*}
\section{Modes and factorization} 
\label{sec:definition}

\subsection{Notation and DIS kinematics in Breit frame} 

In this paper, we work explicitly in the Breit frame, where  a clean geometrical separation of target and current fragmentation exists. To avoid contributions from the resolved photon events we will consider only scattering with large photon virtuality, i.e. $Q = \sqrt{-q^2} \gg 1$ GeV, where $q^{\mu}$ is the four-momentum of the virtual photon. In the Breit frame we have,
\begin{equation}
    q^{\mu}  = \frac{Q}{2}( \bn^{\mu} - n^{\mu} ) = Q (0,0,0,-1)\,,
\end{equation}
where $n^{\mu} \equiv (1,0,0,+1)$ and $\bn^{\mu} \equiv (1,0,0,-1)$.
The proton momentum (up to mass corrections) is:
\begin{equation}
    P^{\mu} \simeq \frac{Q}{2x_B} n^{\mu} = \frac{Q}{2x_B} (1,0,0,+1)\,,
\end{equation} 
with Bjorken $x_B \equiv Q^2 / (2 \,q\cdot P)$. At Born level, the struck quark back-scatters against the proton and has momentum ($x\simeq x_B$):
\begin{equation}
    p_q^{\mu} = x P^{\mu} + q^{\mu} \simeq \frac{Q}{2} \bn^{\mu}\,.
\end{equation}
The fragmentation of the struck quark leads to the formation of a jet-like structure pointing opposite to the beam direction.

In the Breit frame the weight factor, $z_a$, that appears in the definition of the observable in \eq{weigth-DIS} can be written as follows, 
\begin{equation}
    z_a \equiv \frac{P\cdot p_a}{P\cdot (\sum_i p_i)}\Big{\vert}_{P^2 \to 0} = \frac{P \cdot p_a}{P \cdot q}\quad \xrightarrow[\text{frame}]{\text{Breit}} \quad \frac{  p_{a}^{+}}{Q}\,.
\end{equation}
Here and in the rest of this paper we use the standard notation, $p^{+} \equiv n \cdot p$ and $p^{-} \equiv \bar{n} \cdot p$. Note that from momentum conservation  we have 
\begin{equation}\label{eq:sum-rule}
    \sum_{i} z_i = 1\,,
\end{equation}
where the sum extents over all final state particles in the event, excluding the scattered lepton and its QED radiation products. The struck quark fragments, found close the direction of the virtual photon, share the largest fraction of the lightcone momenta $p^+$ and, thus, satisfy $z_i \sim 1$. Then, the constraint in \eq{sum-rule} requires that soft radiation and particles close to the direction of the target hadron are described by parametrically smaller values of the momentum fraction, i.e., $z_i \ll 1$. Therefore, in the back-to-back limit the contribution to \BEEC is dominated by the collinear fragments of the struck quark. We refer to the corresponding modes as ``$\bn$-collinear''. However, from conservation of momentum we have that ``$n$-collinear'' as well as global soft radiation will contribute through recoil, as we demonstrate in the following subsection. Using the light-cone basis where $p^{\mu} = (p^+,p^-,\bmat{p}_{\perp})$, the scaling of the relevant modes is:
\begin{align}
\label{eq:modes}
    n\text{-collinear}&:\;& p_n^{\mu}&\sim Q(\lambda^2,1,\lambda)\;, \nl 
    \bn\text{-collinear}&:\;& p_{\bn}^{\mu}&\sim Q(1,\lambda^2,\lambda)\;, \nl
    \text{soft}&:\;&  p_s^{\mu}&\sim Q(\lambda,\lambda,\lambda) \;.
\end{align}
From the above scaling of momenta we have that the weight factor, $z_i$, for the three relevant modes is 
\begin{align}
     z_{\bn} &\lesssim 1\;,&z_n &\sim \lambda^{2}\;,& z_s&\sim\lambda\;,
\end{align}
while beam remnants with momenta almost parallel to the direction of the proton have a vanishing scaling parameter, $z_{\text{b.r.}} \sim 0$.

\subsection{Factorization} \label{subsec:modes}

As already mentioned above, the $n$-collinear and soft modes will enter the factorization theorem through recoil effects.  To demonstrate that, consider the contribution to the observable of a hadron, $a$, from the $\bn$ sector. As shown in Fig.~\ref{fig:measurment},  the hadron-$a$ is labeled by its scaling factor $z_a$ and its transverse momentum w.r.t. the fragmenting parton $\bmat{k}_{a\perp}$. This particle will contribute to the measurement at $\theta = \theta_{ap}$,  which we can express in terms of $\bmat{q}_{\perp}\equiv \bmat{p}_{a\perp}/z_a$ -- the re-scaled transverse momenta with respect to the proton/beam direction of motion,  
\begin{equation}
    \frac{\pi - \theta_{ap}}{2} = \frac{\bar{\theta}_a}{2} \simeq \frac{| \bmat{q}_{\perp} | }{ Q} \simeq \frac{1}{Q} \Big{|} \frac{\bmat{k}_{a\perp} }{z_a} + \bmat{p}_{n\perp}  - \bmat{p}_{s\perp}   \Big{|} \;.
\end{equation}
It is, therefore, clear now that the EEC is a reweighted TMD distribution summed over all hadrons flavours. Following ref.~\cite{Moult:2018jzp}, we introduce a dimensionless variable in order to simplify the notation, 
\begin{equation}
    \tau = \frac{1+\cos\theta}{2}\;.
\end{equation}
Expanding in the back-to-back limit, $\theta \to \pi$, and for a single hadron of flavour $a$ we have 
\begin{multline}
   \frac{ d\sigma_a }{dx dQ^2 dz d\tau }  = \int d\bmat{q}_{ \perp} \frac{ d\sigma_a }{dx dQ^2 dz d\bmat{q}_{\perp}}  \delta \lp   \tau - \frac{ | \bmat{q}_{ \perp} |^2 }{Q^2} \rp \nl
   \times \lb1+ \mathcal{O}(\tau) \rb \;,
\end{multline}
where we used the shorthand notation $d\sigma_{a} \equiv d\sigma(ep \to e+a+X)$ and we have dropped the subscript $a$ from the variable $z$.

The TMD cross section can be written as usual, in terms TMD parton distribution functions and fragmentation functions:
\begin{multline}
    \frac{ d\sigma_a }{dx dQ^2 dz d\bmat{q}_{\perp}} =  H_{ij} (Q,x,y;\mu) \int \frac{d\bmat{b}}{(2\pi)^2} \exp( -i\;\bmat{b} \cdot \bmat{q}_{\perp})   
    \\[5pt]
   \times  B_{j/P}(x,b;\mu,\nu)   D_{i/a}(z,b;\mu,\nu) S(b;\mu,\nu)\;,
\end{multline}
where $\bmat{b}$ is the Fourier conjugate of $\bmat{q}_\perp $ and $b = |\bmat{b} |$. $B_{j/P}$ is the TMD beam function, $S$ is the soft function, and  $D_{i/a}$ is the fragmentation function for parton $i$ to hadron $a$.

Integrating against the weighing factor and summing over all possible hadrons, we have,
\begin{multline}\label{eq:factorization-BEEC}
 \sum_a   \int_0^1 dz \;z \; \frac{ d\sigma_a }{dx dQ^2 dz d\tau } =  H_{ij} (Q,x,y;\mu) 
 \\[5pt] \times 
 \int d\bmat{q}_\perp \;\delta \lp \tau -  \frac{ | \bmat{q}_\perp |^2 }{ Q^2} \rp \int \frac{ d\bmat{b}}{(2\pi)^2} \exp(-i\; \bmat{b} \cdot  \bmat{q}_\perp)
 \\[5pt] \times
  B_{j/P}(x, b,\mu,\nu)   S(b,\mu,\nu)  J_i (b,\mu,\nu)\;,
\end{multline}
where $J_{q(\bar{q})}$ is the EEC (anti-)quark jet function defined as the weighed sum of TMD fragmentation functions, 
\begin{equation}
\label{eq:jet-def}
    J_i (b;\mu,\nu) \equiv \sum_a \int_0^1 dz\, z  D_{i/a}(z,b;\mu,\nu)\;.
\end{equation}
Note that this is the same jet function that appears in the conventional EEC observable for electron-positron annihilation.  
Through an OPE, the TMDFFs can be expressed in terms of a convolution of short distance matching coefficients, $\mathI_{ij}$, and the collinear fragmentation functions, $d_{i/h}$,\footnote{Note that in the literature there are multiple conventions for the matching coefficients. Here we use the ones from ref.~\cite{Moult:2018jzp}, however other equivalent choices can be made.} 
\begin{multline}
\label{eq:matching}
    D^{\text{OPE}}_{i/a}(z, b;\mu,\nu) =  
    \sum_j \int_z^1 \frac{du}{u}\; \mathI_{ij}\lp\frac{ b }{z/u}, z/u   ,\mu,\nu \rp   \\
   \times d_{j/a}(u,\mu) \lb 1+ \mathcal{O}\lp \frac{\Lambda_{\text{QCD}}^2}{Q^2},\Lambda_{\text{QCD}}^2 b^2\rp \rb \;,
\end{multline}
where we have used the superscript OPE to denote that this is the leading contribution in the expansion and is considered a good approximation of the true TMDs in the perturbative regime where $q_T \gg \Lambda_{\text{QCD}}$. Therefore, for the EEC jet function in the same approximation we have,
\begin{equation}
\label{eq:jet-function}
     J_i^{\text{OPE}} (\bmat{b},\mu,\nu) =
     \\
     \sum_j \int_0^1 dw \; w \; \mathI_{ij}\lp\frac{ b }{w}, w  ;\mu,\nu \rp \;.
\end{equation}
To obtain this, we used the fragmentation function sum rule, 
\begin{equation}
    \sum_a \int_0^1 dz\;z\;d_{i/a}(z;\mu) = 1\;,
\end{equation}
which gives the normalization of the fragmentation functions as number density. 

\subsubsection*{Renormalization group evolution and resummation}

The various elements of the factorized expression have renormalization scale, $\mu$, dependence. The resummed cross-section, where all large logarithms are summed up to a particular logarithmic accuracy, is  obtained by evaluating the elements of factorization at their canonical scales and then using the renormalization group (RG) equations to evolve  them up to a common scale. The RG equation satisfied by each of the relevant functions is, 
\begin{equation}
    \frac{d}{d\ln\mu}G(\mu) = \gamma_G(\mu) G(\mu)\;,
\end{equation}
where $G$ can be any of the functions that appear in the factorization theorem in \eq{factorization-BEEC}. The solution of the RG equation in Laplace space can be written as a product of the evolution kernel $U_G$ and the function $G$, evaluated at some initial scale $\mu_0$, 
\begin{align}
    G(\mu) &= G(\mu_0) \times \exp \lp \int_{\mu_0}^{\mu} d\ln \mu'\, \gamma_G(\mu') \rp \nl
    &= G(\mu_0) \times U_{G}(\mu,\mu_0)\;.
\end{align}
For each of the functions that appear in the factorization theorem, a different choice of the initial scale $\mu_0$ is appropriate. In this work the initial scales of the evolution are chosen such that they minimize large logarithms in the fixed order expansion of the corresponding functions, 
\begin{align}
    \mu_H &= Q\;,&   \mu_S &= \mu_B= \mu_J = \mu_b \equiv \frac{2e^{-\gamma_{E}}}{b}\;.
\end{align}
In our scheme the final scales on the evolution are chosen such that they minimize logarithms in the beam and jet functions, $\mu= \mu_b$. In addition to the renormalization evolution, the soft, beam, and jet functions  also satisfy rapidity evolution. Here, we  consider the evolution of the soft function alone, 
\begin{align}
    S(b,\mu,\nu) = S(b,\mu,\nu_S) \times \exp \lp \int_{\nu_S}^{\nu} d\ln \nu'\, \gamma_R(\mu, \mu_b) \rp \;,
\end{align} 
and choose the common rapidity scale, $\nu$ such that it minimizes the rapidity logs in the beam and jet function:   
\begin{equation}
     \nu= \nu_B = \nu_J = Q\;.
\end{equation}
The anomalous dimension $\gamma_R$ is the soft function rapidity anomalous dimension. Then, the RG evolved cross section reads
\begin{multline}{\label{eq:resum}}
    \frac{d\sigma}{dx dQ^2 d\tau }= H_{ij} (Q,x,y;\mu_H) 
 \int d\bmat{q}_\perp \;\delta \lp \tau -  \frac{ | \bmat{q}_\perp |^2 }{ Q^2} \rp \\[5pt] 
 \int \frac{ d\bmat{b}}{(2\pi)^2}  
 \exp(-i\; \bmat{b} \cdot  \bmat{q}_\perp)\; B_{j/P}(x, b,\mu,\nu)\; S(b,\mu_S,\nu_S) \\[5pt] 
 \times  \;  J_i (b,\mu,\nu)\;
        R(b,\mu, \nu) \, . 
\end{multline}
Here, $R$ is the combined evolution kernel for both renormalization evolution and rapidity evolution,
\begin{equation}
    R(b,\mu, \nu) = \exp\lp
    \int_{\mu_H}^{\mu} \frac{d\mu'}{\mu'} \gamma_H(\mu')  + \int_{\nu_S}^{\nu} \frac{d\nu'}{\nu'} \gamma_R(\mu, \mu_b) \rp\;,
\end{equation}
where $\gamma_H$ and $\gamma_S$ are the anomalous dimensions of the hard and soft functions. 

Further details on evolution equations and anomalous dimensions can be found in ref.~\cite{Li:2020bub} and references therein.

\subsection{Hadronization  and non-perturbative corrections}
\label{sec:hadronization}
The result in eq.(\ref{eq:jet-def}) contains the same jet function found in the conventional EEC observable in refs.~\cite{Moult:2018jzp,Luo:2019hmp} and can be expressed in terms of the perturbatively calculable matching coefficients after OPE, as done in eq.(\ref{eq:jet-function}). However, hadronization corrections are expected to be important, particularly  at the relatively small values of $Q$  anticipated at the EIC. The fact that \BEEC factorization involves the universal back-to-back TMD soft function allows us to consider hadronization and non-perturbative corrections in a universal framework applicable to conventional TMD observables.

There are two sources of non-perturbative corrections for TMDs: a) the corrections to the rapidity anomalous dimension; and b) the contribution to the TMD matrix elements. For the soft rapidity anomalous dimension the implementation of the non-perturbative model is done as in conventional TMD observables at the level of evolution,
\begin{equation}
     R(b,\mu,\nu)  \to  R(b, \mu,\nu) \times \exp\lp g_K(b) \ln\frac{\nu}{\nu_S} \rp\;,
\end{equation}
where  $g_K(b)$ is the model function for the non-perturbative component of the rapidity anomalous dimension. Note that since the EEC jet function satisfies the same rapidity and renormalization group evolution as TMDFFs, this also implies the same hadronization model for the rapidity anomalous dimension. For the hadronization model of the EEC jet function we can assume a generic multiplicative ansatz. The final result reads,
\begin{equation}\label{eq:ansatz}
   \sqrt{S} J_i( b, \mu_0,\nu_0) = \sqrt{S^{\text{pert.}}} J^{\text{OPE}}_i(b,\mu_0,\nu_0) \, j_{i}(b) \;,
\end{equation}
where $S^{\text{pert.}}$ is the perturbative expression for the soft function and  $j_{i}(b)$ is the multiplicative ansatz for hadronization effects in the EEC jet function. The scales $\mu_0$ and $\nu_0$ are  arbitrary scales in the soft and jet functions.  Since this is the same jet function that appears in the standard EEC observable  one can extract the model function  $j_{i}(b)$ by fitting to experimental data from $e^+ e^-$ or in a simultaneous fit extraction of jet and TMDPDF from $ep$ colliders.

Alternatively, we can relate the model function $j_{i}(b)$ to the standard TMDFFs as follows,
\begin{multline}\label{eq:model-jet}
     j_{i}(b) = \lb J^{\text{OPE}}_i(b,\mu_0,\nu_0) \rb^{-1}  \sum_a \int_0^1 du  dy \,(uy) \; \\[5pt] \mathI_{ij}\lp\frac{ b }{u}, u  ;\mu_0,\nu_0 \rp \; d_{i/a} (y,\mu_0)\; D_{i/a}^{\text{NP}}(uy, b) \;,
\end{multline}
where the non-perturbative model function $D_{i/a}^{\text{NP}}(z, b)$ is defined in the context of TMDFFs,
\begin{equation}
\label{eq:model-FF}
    \sqrt{S} D_{i/a}(z,b;\mu_0,\nu_0) = 
    \sqrt{S^{\text{pert.}}} D^{\text{OPE}}_{i/a}(z, b;\mu_0,\nu_0)D_{i/a}^{\text{NP}}(z, b)\;.
\end{equation}
Therefore, given a specific model for TMD fragmentation we can explicitly evaluate the model function $j_{i}(b)$. This is the approach we take here. We use the approach given above to calculate the jet function from past extractions of the TMDFF. Relatively recent extractions of the TMDFFs can be found in refs.~\cite{Su:2014wpa,Bacchetta:2017gcc,Scimemi:2019cmh}. The non-perturbative model for the TMDPDF can be implemented the same way as is done for in the case of TMD measurements in SIDIS and Drell-Yan processes,
\begin{equation}
\label{eq:model-PDF}
   \sqrt{S} B_{i/P}(x,b;\mu_0,\nu_0) = \sqrt{S^{\text{pert.}}} B_{i/P}^{\text{OPE}}(x,b;\mu_0,\nu_0)
     f_{i/P}^{\text{NP}}(x,b)\;.
\end{equation}
In this work we will be assuming  a simplified  model that is independent of the Bjorken $x$:
\begin{equation}
    f_{i/ P}^{\text{NP}}(x,b)\Big{\vert}_{\text{simplified}} =  f_{i/P }^{\text{NP}}(b)\;.
\end{equation}
Thus, combining all elements at the level of the cross section we can collect all non-perturbative contributions in a single function, $F_{i/P}^{\rm NP}(b)$,
\begin{multline}{\label{eq:resum}}
    \frac{d\sigma}{dx dQ^2 d\tau }= H_{ij} (Q,x,y;\mu_H) 
 \int d\bmat{q}_\perp \;\delta \lp \tau -  \frac{ | \bmat{q}_\perp |^2 }{ Q^2} \rp \\[5pt] 
 \int \frac{ d\bmat{b}}{(2\pi)^2}  
 \exp(-i\; \bmat{b} \cdot  \bmat{q}_\perp)\; B_{j/P}^{\text{OPE}}(x, b,\mu,\nu)  
   \; S^{\text{pert.}}(b,\mu_S,\nu_S) \\[5pt] 
 \; \times  J_i^{\text{OPE}} (b,\mu,\nu)\;
        R(b, \mu, \nu )\;F_{i/P}^{\rm NP}(b) \, , 
\end{multline}
where 
\begin{equation}
    F^{\text{NP}}_{i/P}(b) = f_{i/ P}^{\text{NP}}(b)\; \;j_i^{\text{NP}}(b)\;\exp\lp g_K(b) \ln\frac{\nu}{\nu_S} \rp\;,
\end{equation}
The non-perturbative corrections in this work are implemented  following the model and parameters in refs.~\cite{Su:2014wpa,Prokudin:2015ysa}, 
\begin{multline} \label{eq:np}
    F_{i/P}^{\rm NP}(b) \sim j_i(b)\times \exp \lb -0.106\  b^2 \\
    - 0.42\ln \lp 1+ \frac{b^2}{b^2_{\max}} \rp\ln \frac{\nu}{\nu_S} \rb \, , 
\end{multline}
where the last term in the exponent corresponds to the non-perturbative rapidity evolution model for $g_K(b) =  - 0.42 \ln (1+b^2/b_{\max}^2)$, with $b_{\max} = 1.5\; \rm{GeV}^{-1}$. For the EEC jet function model $j_i(b)$ we will consider a LO approximation based on \eq{model-jet}, 
\begin{equation}\label{eq:LO-j}
    j_{i}(b) \Big{\vert}_{\text{LO}} = \sum_a \int_0^1dy\,y\,d_{i/a}(y,1 \,\text{GeV})  D_{i/a}^{\text{NP}}(y, b)\,.
\end{equation}
Note that for the trivial choice $D_{i/a}^{\text{NP}} = 1$ (i.e., no non-perturbative effects in the TMDFF) we have from momentum sum rules that also $j_{i}(b) = 1$.  For this analysis we use the DSS collinear fragmentation functions~\cite{deFlorian:2007ekg,deFlorian:2007aj} and for the TMDFF model we have, 
\begin{equation}
    D_{i/a}^{\text{NP}}(y, b) = \exp\lp -0.042 \frac{b^2}{y^2}\rp\;.
\end{equation}
Performing the sum over $a \in \{\pi^{+/-/0},\;p/n,\;K^{+/-/0}\}$ and the integrating over the momentum fraction $y$ in \eq{LO-j} we find, 
\begin{equation}
    j_{i}(b) = \exp\lp -0.59 b-0.03b^2 \rp\,,
\end{equation}
for $i\in \{u,d,s,\bar{u},\bar{d},\bar{s}\}$ with very small flavour dependence which in this study we will ignore for simplicity.

In this section we discussed how the non-perturbative corrections are implemented in our work. Our approach is phenomenologically motivated and it aligns with what is traditionally used for TMD extractions from semi-inclusive DIS. Nonetheless, there are  other ways to take into account non-perturbative effects in EEC. A detailed  field theoretic treatment of non-perturbative corrections to EEC in $e^+e^-$ was presented in ref.~\cite{Dokshitzer:1999sh}. We find that the linear in $b$ terms in the non-perturbative component of the jet function are consistent with the NP-model introduced in~\cite{Dokshitzer:1999sh}.  The  question  of power corrections and non-perturbative effects in a field theoretic manner is a rather interesting and important subject, but beyond the scope of this paper. 
\begin{figure*}[t!]
    \centering
    \includegraphics[width=0.48\textwidth]{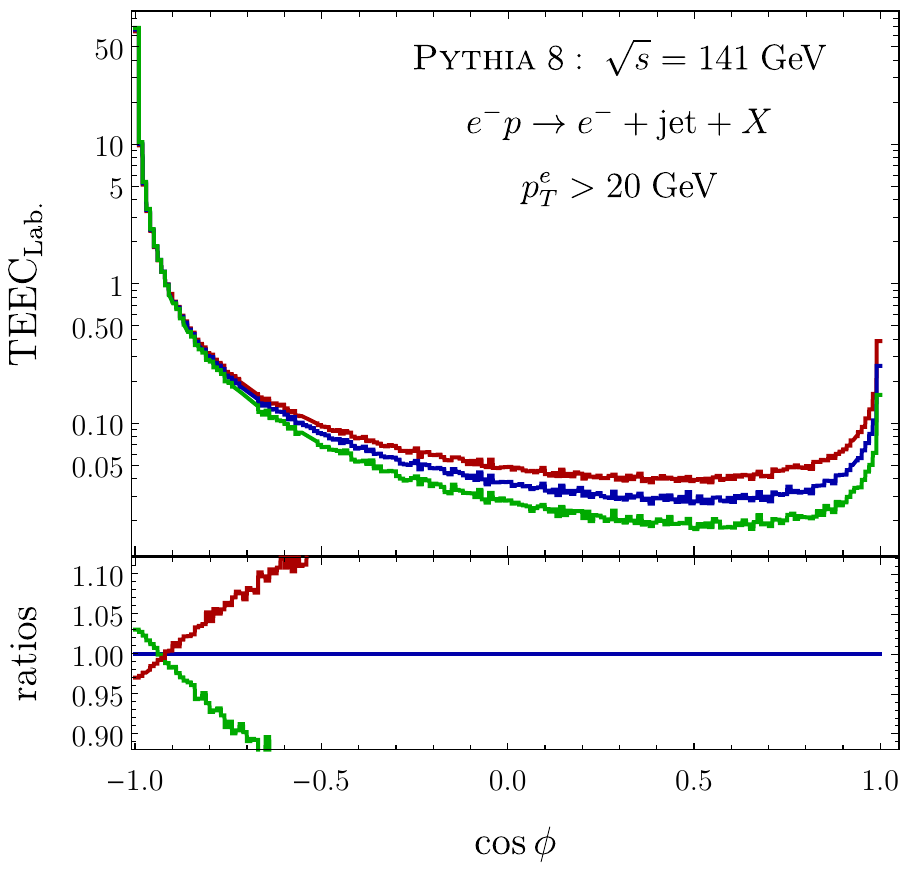}\;\;\hspace{0.4cm}
    \includegraphics[width=0.48\textwidth]{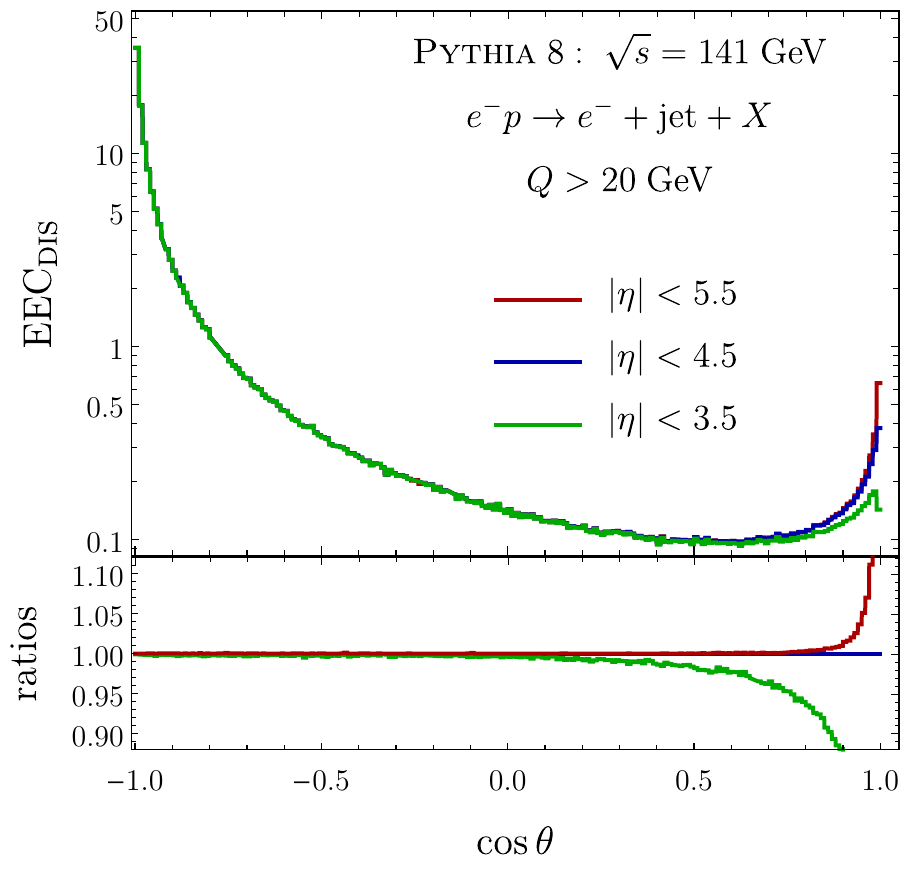}
    \vspace{-0.5cm}
    \caption{TEEC$_{\rm Lab}$ (left) and EEC$_{\rm DIS}$ (right) distributions from PYTHIA 8 with different rapidity cuts in the lab frame. For TEEC$_{\rm Lab}$ we consider a cut on the scattered lepton transverse momentum $p_T^e > 20$ GeV and for EEC$_{\rm DIS}$ we consider cut on the photon virtuality $Q > 20$ GeV. For both case we consider cut on $y = 2 P\cdot q /s$: $0.01 < y < 0.95$.}
    \label{fig:pythia}
\end{figure*}
\section{Numerics and comparison with Monte-Carlo simulations} \label{sec:numerics}

In this section we  present numerical results for the EEC$_{\rm DIS}$ distributions at the future Electron-Ion Collider (EIC)~\cite{Aschenauer:2014cki}. In particular, we  consider the following  beam energies: 18 GeV electrons on 275 GeV protons, which corresponds to the center-of-mass energy $\sqrt{s}\approx 141$ GeV. For all the calculations we select events with $Q>20$ GeV and  use the PDF4LHC15\_nnlo\_mc PDF sets~\cite{Butterworth:2015oua}  with the associated strong coupling provided by  {\sc \small Lhapdf 6}~\cite{Buckley:2014ana}. 

We present the TEEC$_{\rm Lab}$~\cite{Li:2020bub} and EEC$_{\rm DIS}$ distributions predicted by \textsc{Pythia} 8~\cite{Sjostrand:2007gs,Sjostrand:2014zea} in Fig.~\ref{fig:pythia}. The  red,  blue, and  green lines represent the results with pseudorapidity cuts $|\eta|<5.5$, $|\eta|<4.5$, and $|\eta|<3.5$ in the lab frame, respectively, which imitates detector limitations in the backward and forward regions. In order to compare the results with different pseudorapidity cuts,  all the distributions in Fig.~\ref{fig:pythia} are normalized by the sum of weights of all entries relative to the same quantity for $|\eta|<5.5$.   Because TEEC$_{\text{Lab}}$ measures the correlation between hadrons and the final state lepton in the lab frame, pseudorapidity cuts have an impact on the full $\cos\phi$ range, as shown in left panel of Fig.~\ref{fig:pythia}. EEC$_{\text{DIS}}$ is defined as the correlation between the final state hadrons and incoming proton in the Breit frame, and the pseudorapidity cuts only remove particles in the forward region where the weighted cross section is small. In the backward region ($\cos\theta \to -1$) the EEC$_{\rm DIS}$ distribution is independent on the pseudorapidity cuts.

\begin{figure}[t!]
    \centering
    \includegraphics[width=0.48\textwidth]{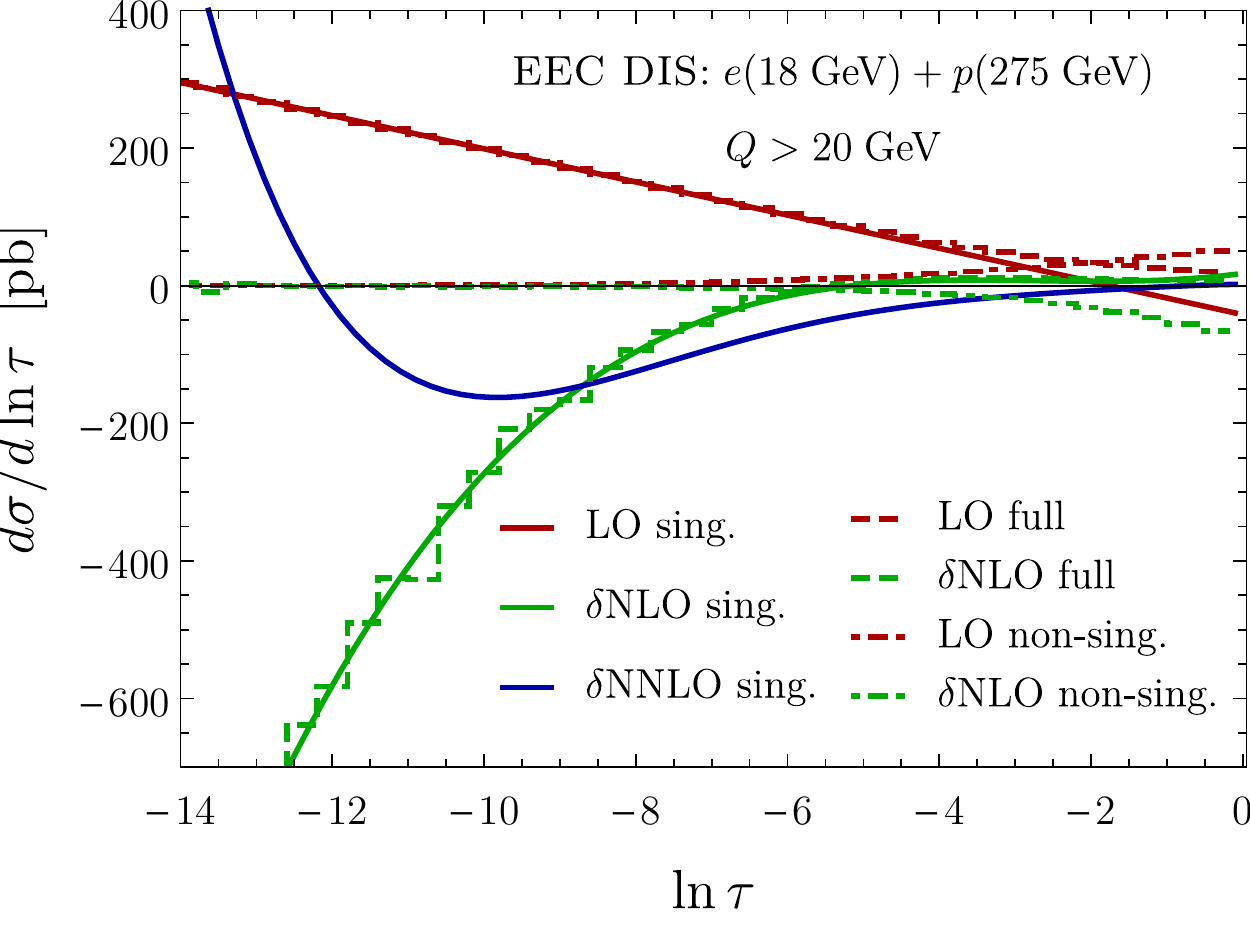}
    \vspace{-0.6cm}
    \caption{Fixed-order $\ln\tau$ distributions in the $\tau \to 1$ limit.  The solid lines represent the singular distribution predictions of SCET. The dashed lines are the fixed order QCD results. The dash-dotted lines are  power corrections.  }
    \label{fig:log_FO}
\end{figure}

Fig.~\ref{fig:log_FO} shows the fixed order $\ln\tau$ distributions in QCD compared against the singular distributions derived from the fixed order expansion of the factorized cross section in \eq{factorization-BEEC}. The factorization and renormalization scales are set to  $Q$. 
The non-trivial contribution to \BEEC starts from two-jet production in DIS, which is considered as LO . 
The singular distributions are provided  up to NNLO and shown with solid lines; their $\tau \to 0$ behavior is driven by the logarithmic terms. The fixed order QCD predictions are shown with the dashed lines, and are calculated using \textsc{NLOJet}\small{++}~\cite{Nagy:2001xb,Nagy:2005gn}. The non-singular contributions are defined as 
\begin{equation}
\frac{d\sigma}{d\ln\tau}\Big{|}_{\text{non-sing.}} \equiv \frac{d\sigma}{d\ln\tau}\Big{|}_{\text{QCD}}-\frac{d\sigma}{d\ln\tau}\Big{|}_{\text{sing.}}\;,
\end{equation}
and are represented by the dash-dotted lines. The non-singular results correspond to the power corrections to the factorized cross section which are of $\mathcal{O}(\tau)$ or higher and vanish in the limit $\tau \to 0$. This, equivalently, implies that the singular behavior of the full QCD calculation is reproduced by the singular distributions in the same limit. The purpose of this comparison is twofold. First, it is a consistency check of our approach and our calculations, and, second, it is an indicator for the size of power corrections as a function of $\tau$. In particular, we find that large power corrections are observed for $\tau \to 1$ where the TMD factorization is not valid.

\begin{figure}[t!]
    \centering
    \includegraphics[width=0.48\textwidth]{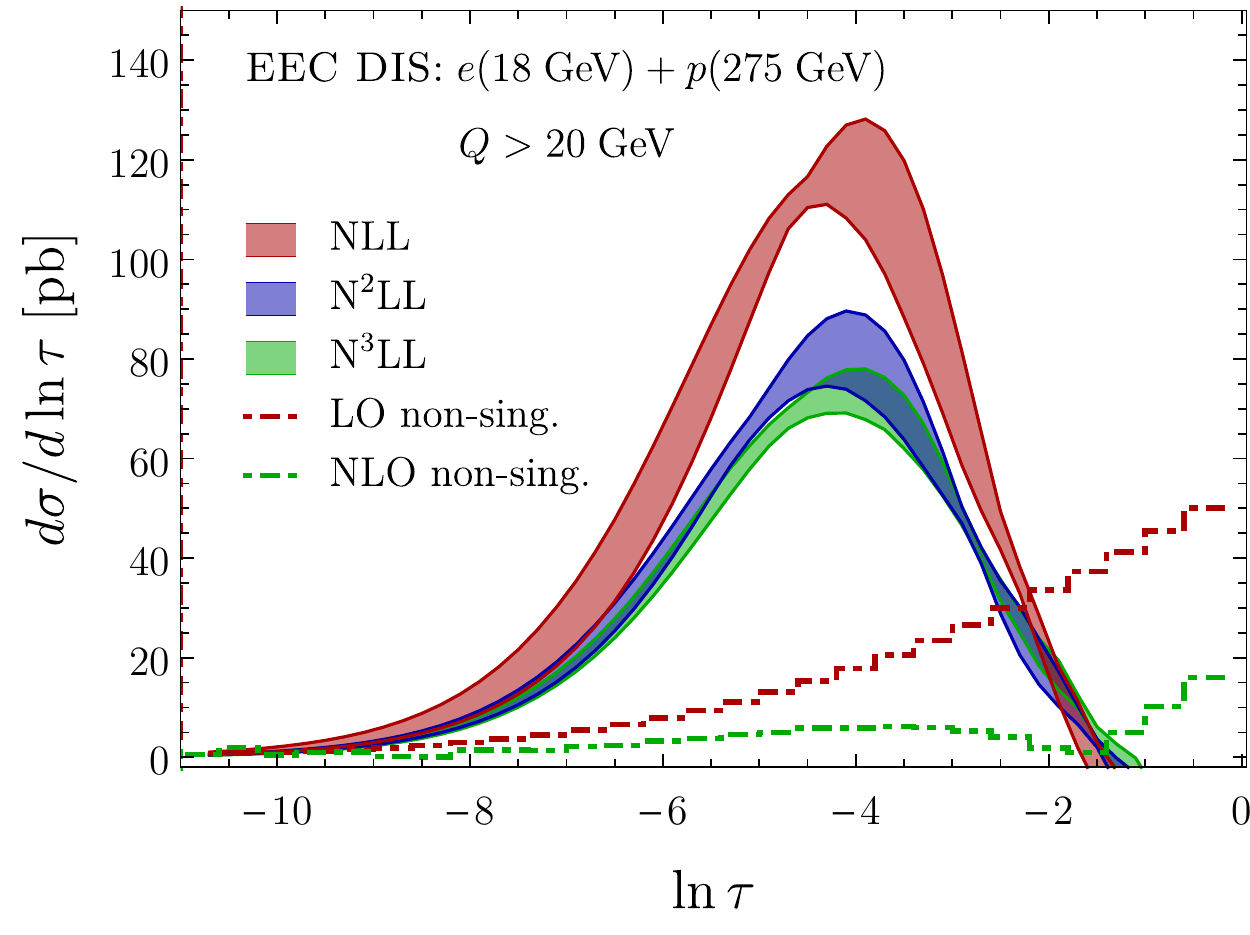}
    \vspace{-0.6cm}
    \caption{Resummed $\ln \tau $ distributions for EEC. The dark red, dark blue, and dark green bands correspond to NLL, NNLL, and NNNLL distribution. The dot-dashed lines are the LO non-singular distributions (dark red) and the absolute value of NLO non-singular ones (dark green). }
    \label{fig:resum}
\end{figure}

\begin{figure*}[t!]
    \centering
    \includegraphics[width=0.48\textwidth]{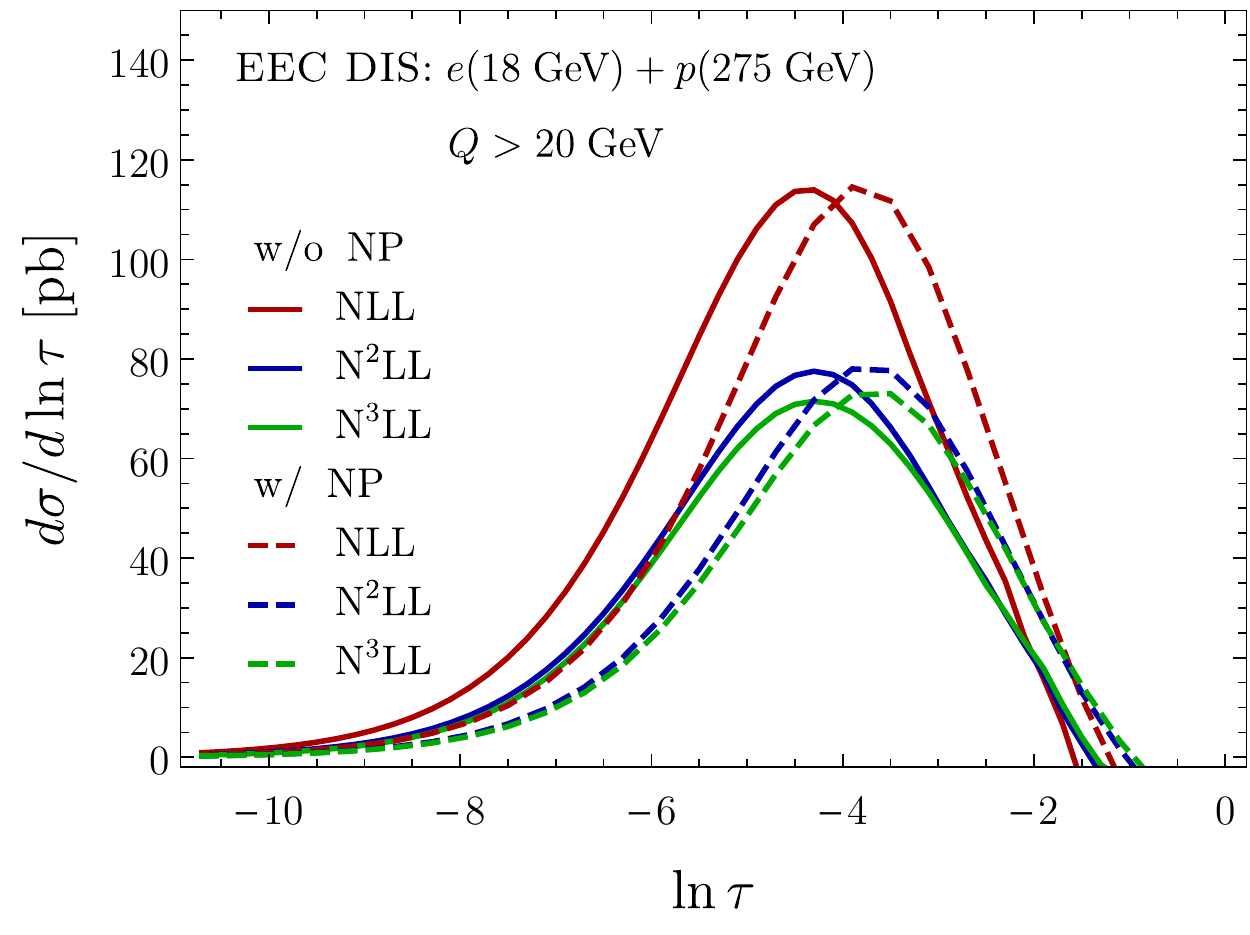}\;\;\hspace{0.4cm}
    \includegraphics[width=0.48\textwidth]{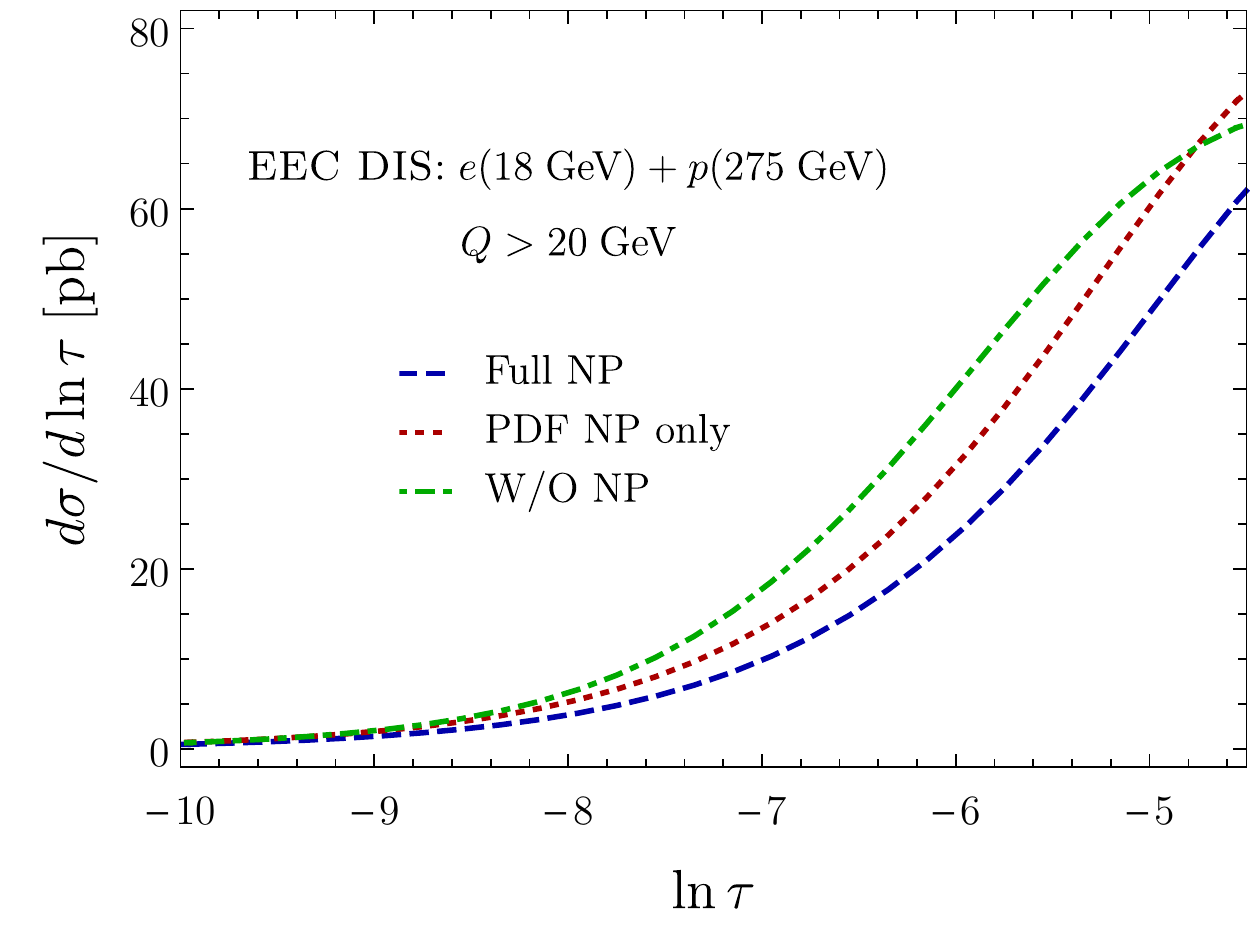}
    \vspace{-0.2cm}
    \caption{Left: Resummed $\ln\tau$ distributions without (solid) and with non-perturbative factor. The color scheme is the same as  in Fig.~\ref{fig:resum}.  Right:  Comparing the effect of the hadronization effects and TMDPDF effects in the small transverse momentum region for N$^3$LL only.}
    \label{fig:resumNP}
\end{figure*}

\begin{figure*}
    \centering
    \includegraphics[width=0.48\textwidth]{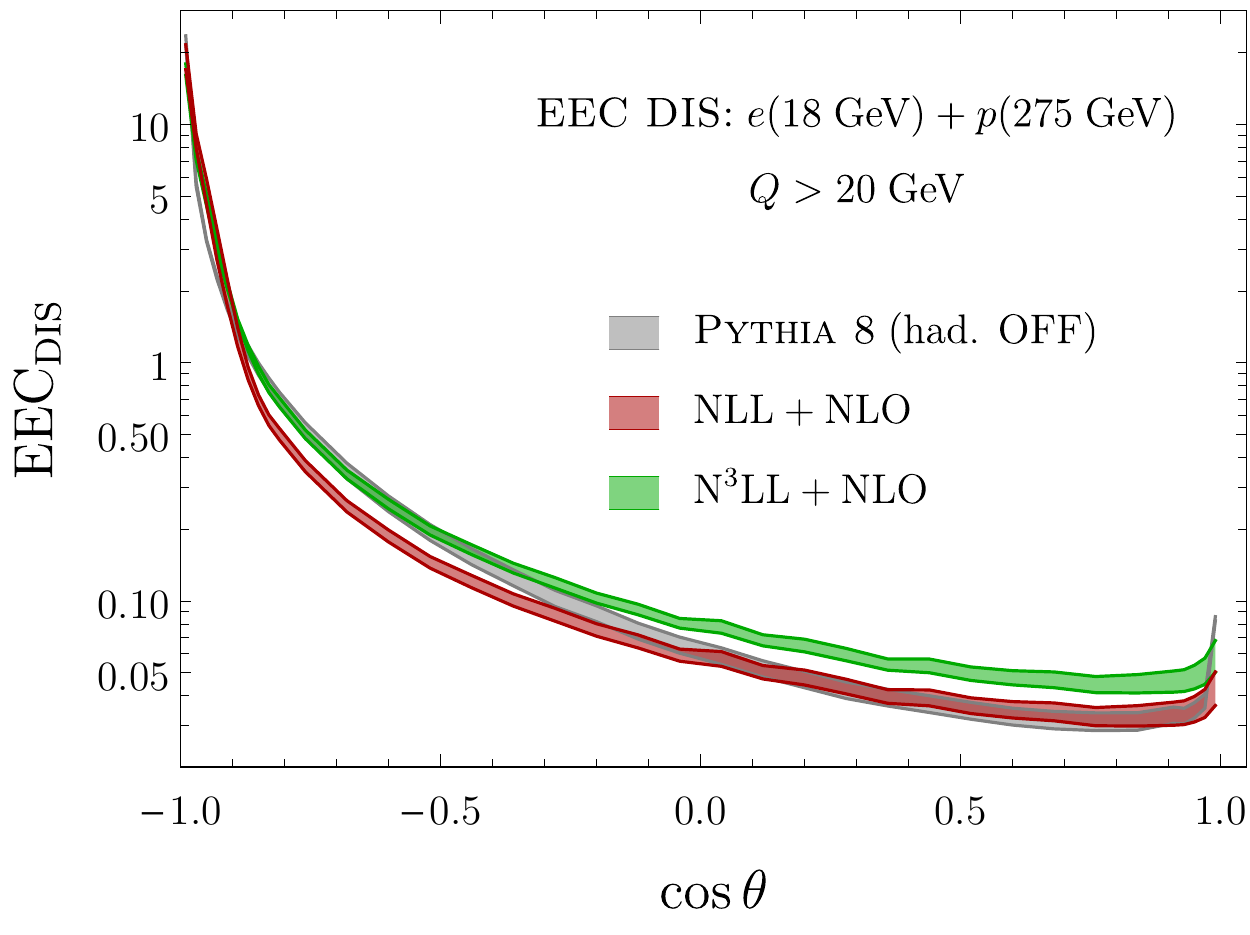}\;\;\hspace{0.4cm}
    \includegraphics[width=0.48\textwidth]{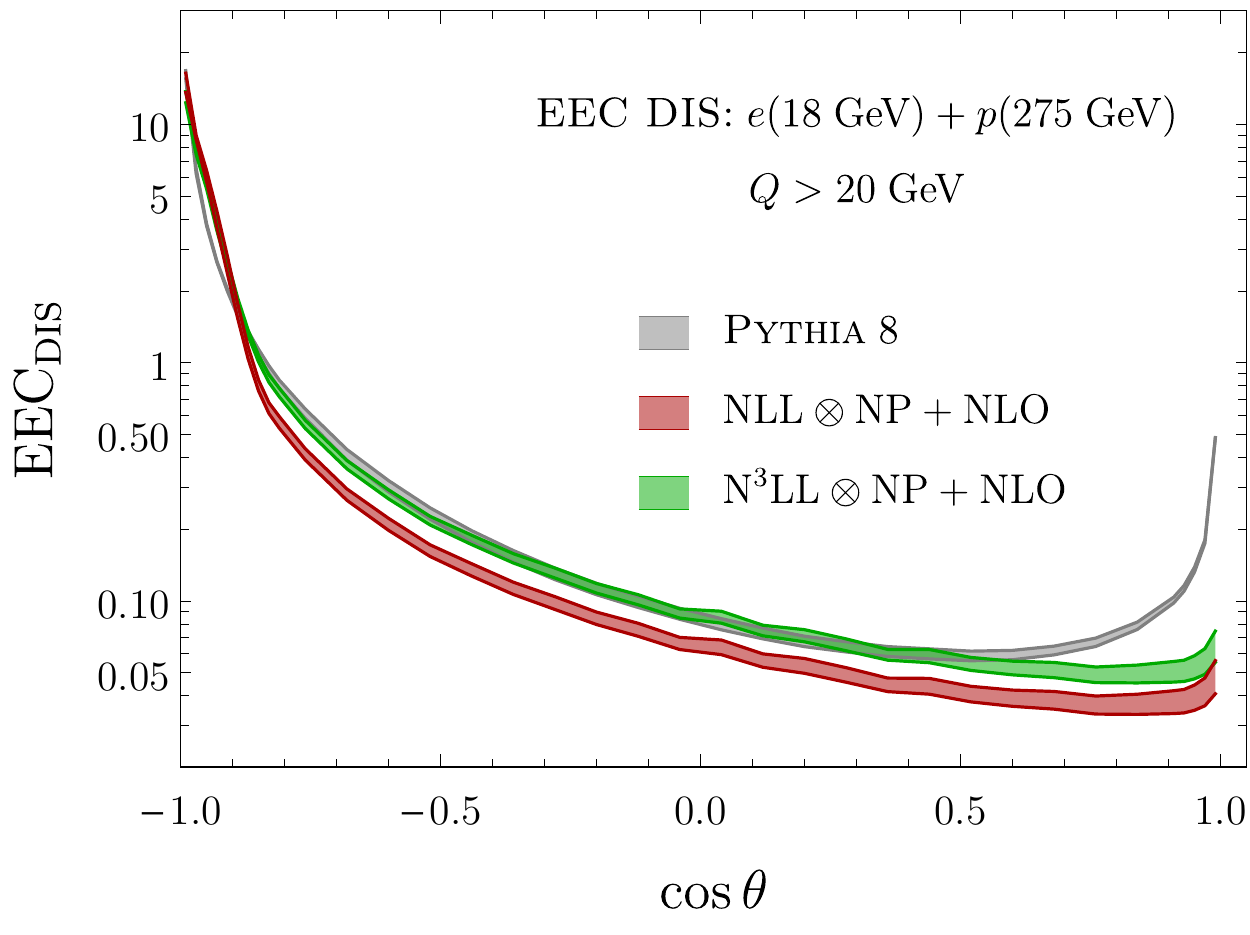}
    \vspace{-0.2cm}
    \caption{Comparison of $\cos\theta$ distributions between the SCET predictions and PYTHIA simulations. The dark red and dark green bands are the NLL+NLO and N$^3$LL+NLO. The gray bands are from PYTHIA 8 simulations with the default settings for uncertainty bands. }
    \label{fig:res_fo}
\end{figure*}

In the $\tau \to 0$ limit the logarithmic enhancements can spoil the convergence of the perturbative expansion. This is apparent from the relative size of the various terms  in Fig.~\ref{fig:log_FO}. Therefore, resummation of these logarithms to all orders in the strong coupling is necessary for reliable predictions and comparison with experimental data. The resummed cross section can be evaluated by evolving the hard, soft, beam, and jet functions in \eq{factorization-BEEC} from their canonical scales to common rapidity and renormalization scales, $\nu$ and $\mu$ respectively.  As discussed in section~\ref{subsec:modes},
we chose to evolve the hard function to the common soft and jet renormalization scales: $\mu = \mu_S =  \mu_B =\mu_J$. For the rapidity evolution we evolve the soft function up to the common jet and beam rapidity scale by making the choice $\nu = \nu_B = \nu_J$. In our scale choice, in order to avoid the Landau pole, we adopt the conventional scheme, 
\begin{equation}
    \nu_S = \mu_S=\mu_B = \mu_J: \quad b\to b^{*}  = \frac{b}{\sqrt{1+b^2/b_{\max}^2}}\;,
\end{equation}
where $b_{\max}$ is fixed as discuss above \eq{np}. The scale uncertainties are evaluated by varying $\mu$, $\mu_H$, $\nu$ and $\nu_H$ by a factor of two independently.  Fig.~\ref{fig:resum} shows resummed distributions over $\ln \tau$ in the infrared region. We observe large corrections from NLL to N$^2$LL and a good perturbative convergence from N$^2$LL to N$^3$LL. Furthermore we find that the scale uncertainties are significantly reduced for the N$^3$LL compared to the lower accuracy distributions. For comparison, in Fig.~\ref{fig:resum} we also show the LO and NLO non-singular distributions. We find that the power corrections become numerically relevant  for $\ln \tau \gtrsim -4$. 

In comparison with the equivalent TEEC$_{\rm Lab}$ distributions, i.e. Fig. 3 of ref.~\cite{Li:2020bub}, the peaks of EEC$_{\rm DIS}$ distributions are at larger $\tau$ values, which corresponds to a larger energy scale. This suggests that the EEC$_{\rm DIS}$ observable has a more stable and reliable perturbative behavior at the bulk of the distribution. Also, compared to TEEC$_{\rm Lab}$, we expect smaller non-perturbative corrections since, in principle, the hadronization effects will suppress the cross section in the region $q_T \sim 0$. For the non-perturbative models discussed in  section~\ref{sec:hadronization} the numerical results are present in Fig.~\ref{fig:resumNP} (left panel). As anticipated, for EEC$_{\rm DIS}$, non-perturbative corrections shift the cross sections to larger $\tau$.  In the right panel the same figure we focus on the very small transverse momentum region in which sensitivity to TMDPDF is stronger. In this region we show the effect of the different non-perturbative components. We note that the hadronization effects are as important in magnitude as the TMDPDF, at least for the extractions from ~\cite{Su:2014wpa,Prokudin:2015ysa} which we used in this paper.

The final EEC$_{\rm DIS}$ distributions are presented in Fig.~\ref{fig:res_fo}, where we matched the NLL and N$^3$LL resummed distributions to the QCD NLO ones in the $\tau\to1$ region. The left panel shows the comparison between our calculations without non-perturbative effects to \textsc{Pythia 8} simulations run without a hadronization modeling. The right panel presents the equivalent distributions including non-perturbative effects and hadronization. For $-0.8 < \cos\theta$ the distributions are described by the fixed order results and for $\cos\theta<-0.95$  by the sum of resummed and non-singular power corrections. In region $ -0.95 <\cos\theta  < -0.8$ we impose our matching scheme where the cross section smoothly transitions from the resummed and the fixed order cross section and is defined as
\begin{align}
    \frac{d\sigma}{d\cos\theta} \Big{|}_{-0.95< \cos\theta < -0.8} =& (1-f^2) \frac{d\sigma}{d\cos\theta} \Big{|}_{\text{QCD}}
    \nonumber \\& \hspace{-2cm}
    +f^2 \left( \frac{d\sigma}{d\cos\theta} \Big{|}_{\text{non-sing.}} +\frac{d\sigma}{d\cos\theta} \Big{|}_{\text{Res.}} \right)\;,
\end{align}
where 
\begin{equation}
    f=\frac{1}{2}\lb \cos\left(\frac{\cos\theta+0.95}{0.15}\pi\right)+1 \rb \;.
\end{equation}
A similar matching procedure and detailed discussion about matching can be found in ref.~\cite{Becher:2019bnm}.  
The difference between NLL+NLO and N$^3$LL+NLO in the region $\cos\theta > -0.8$ is purely  from the total normalization, which is dominated by the resummation region. 

For large $\cos\theta$, where the cross section is dominated by the fixed order result, the theoretical  uncertainties are estimated by varying the renormalization and factorization scales ($\mu_r$ and $\mu_f$ respectively) by a factor of 2 and 1/2: $\mu_r=\mu_f=\kappa Q$ with $\kappa=(0.5,1,2)$.

By comparing the \textsc{Pythia} distributions (gray band) in the left and right panels of Fig.~\ref{fig:res_fo}  we see that  hadronization effects, as implemented in \textsc{Pythia 8}, enhance the distribution for large $\cos\theta$. In the collinear limit, $\theta \to 0$, the distribution from hadronic-level \textsc{Pythia} increases  in comparison to the partonic-level and in comparison with our NLO predictions. Note, however, that this region is sensitive to the rapidity cutoffs and thus comparison to experiment requires a different formalism to incorporate these effects. Such a formalism  will have to resum logarithmic enchantments from the rapidity cutoff, see for example refs.~\cite{Hornig:2017pud,Kang:2018agv,Michel:2018hui}. 
\section{Correlations with subsets of hadrons} \label{sec:subsets}
In this section we discuss an EEC-like observable, but this time considering only a subset of hadrons. There are various reasons why one might want to introduce  such a modification and at the end of this section we give two possibilities. In the discussion that follows, we denote the subset of hadrons with $\mathbb{S}$. Taking the $\tau$-differential and  $z$-weighed cross section but only summing over a subset of all hadrons we have~\footnote{For EEC$_{\rm DIS}$, there is no collinear singularity in the forward region. In this section we focus on the back-to-back limit. },
\begin{equation}
  \mathcal{O}_{\mathbb{S}}  = \sum_{a \in \mathbb{S} }   \int_0^1 dz \;z \; \frac{ d\sigma_a }{dx dQ^2 dz d\tau } \;.
\end{equation}
In the back-to-back limit the factorization theorem formulation for this observable follows exactly the same steps as for \BEEC and the final result is given by eq.~(\ref{eq:factorization-BEEC}) with the replacement of the EEC jet function with the appropriate \emph{subset jet function}, $\mathcal{G}_{i\to \mathbb{S}}$:
\begin{equation}
    J_i (b,\mu,\nu) \to \mathcal{G}_{i\to \mathbb{S}} (b,\mu,\nu) = \sum_{a \in \mathbb{S} } \int_0^1 dz\, z  D_{i/a}(z,\bmat{b};\mu,\nu)\;.
\end{equation}
Beyond the dependence on the initiating parton $i$, the subset  jet function (SJF)   also depends on the subset of hadrons $\mathbb{S}$ and, thus,  is a fundamentally non-perturbative function (similar to fragmentation functions and PDFs).  Nonetheless, similarly to the EEC jet function we can employ the OPE and at leading order match the TMD FF onto the collinear FFs. Doing so we get, 
\begin{equation}\label{eq:SJF}
\mathcal{G}_{i \to \mathbb{S}}^{\text{OPE}} (\bmat{b},\mu,\nu) = \sum_j \mathcal{F}_{j \to \mathbb{S}} (\mu) \int_0^1 dw \; w \; \mathI_{ij}\lp\frac{ \bmat{b} }{w}, w  ;\mu,\nu \rp \;,
\end{equation}
where  $ \mathcal{F}_{j \to \mathbb{S}}$ is interpreted as the average fraction the momentum of the fragmenting parton $j$ carried by the hadrons included in $\mathbb{S}$,
\begin{equation}
     \mathcal{F}_{j \to \mathbb{S}}(\mu) = \sum_{a \in \mathbb{S}} \int_0^1 dz\,z\,d_{i\to a}(z,\mu)\;.
\end{equation}
From the fragmentation function sum rules we also have that if $\mathbb{S} = $ \emph{all hadrons} then $\mathcal{F}_{j \to \mathbb{S}} = 1$ and we retrieve the EEC jet function. In-contrast to the EEC jet function, the SJF in eq.~(\ref{eq:SJF}) cannot be expressed only in terms of the perturbatively calculable matching coefficients since it depends explicitly on the subset $\mathbb{S}$. However, in the OPE limit this dependence is rather simple and enter through the  fraction  $\mathcal{F}_{j\to \mathbb{S}}$. Beyond this limit, further non-perturbative corrections can be implemented through a model in a way similar to  the one for the EEC jet function in eq.(\ref{eq:ansatz}):
\begin{equation}\label{eq:ansatz}
    \sqrt{S} \mathcal{G}_{i\to \mathbb{S}}(\bmat{b}, \mu,\nu) = \sqrt{S^{\text{pert}}} R(\bmat{b},\mu, \nu) \, \mathcal{G}^{\text{OPE}}_{i \to \mathbb{S}}(\bmat{b},\mu_0,\nu_0) \, g_{i \to \mathbb{S}}(\bmat{b})\;,
\end{equation}
where 
\begin{equation}\label{eq:model}
     g_{i \to \mathbb{S}}(\bmat{b}) = \lb \mathcal{G}^{\text{OPE}}_{i\to \mathbb{S}}(\bmat{b},\mu_0,\nu_0) \rb^{-1}  \sum_{a \in \mathbb{S}} \int_0^1 dz\, z  D_{i/a}(z,\bmat{b};\mu_0,\nu_0)\;,
\end{equation}
is the non-perturbative ansatz  for the SJF to be determined from the flavor sensitive TMD FFs or directly measured from experiment.

The subset of hadrons $\mathbb{S}$ can be chosen accordingly in order to accommodate the experimental and phenomenological needs. Two cases we find particularly interesting:
\begin{itemize}
    \item $ \mathbb{S} = \mathbb{C}$: where $\mathbb{C}$ is the subset of all charged particles. This case is important since often experimental measurements consider only charged particles for which the energy resolution is much better, yielding  smaller experimental uncertainties. To apply our formalism in the back-to-back limit one needs the fraction  $\mathcal{F}_{i \to \mathbb{C}}$ which can be obtained from extraction of charge hadron fragmentation functions~\cite{deFlorian:2007ekg}. Considering only charged particles will most definitely have an effect in the weighed cross section, however,  is unclear how much  this will affect the \BEEC since it is a normalized observable. 
    \item $ \mathbb{S} = h $: where $h$ is an identified hadron. This case is particularly interesting since it allows us to probe the  flavor of the incoming parton by appropriately choosing the final state hadron $h$. This is similar to what has been done in TMD observables~\cite{Signori:2013mda}. Obviously,  the non-perturbative number $\mathcal{F}_{i \to h}$ can be obtain by integrating the single hadron collinear FF, $D_{i \to h}$.
\end{itemize}
In this work we do not explore any numerical implementations of this modified EEC-like observable but we foresee those to be relevant for future studies when comparing to experimental data.    

\section{Conclusion} \label{sec:concl}
In this paper, we introduced a new definition of energy-energy-correlator (EEC) suitable for Breit frame studies of the DIS process, where the opening angle $\theta$ between final state particles and the incoming proton is measured and the cross section is weighted by the four-momentum product of final state particles and incoming proton.
As a consequence of this definition, the contributions that arise from wide angle soft particles and initial-state radiation close to the beam direction, are suppressed. The usual transverse momentum factorization can be applied in the back-to-back limit ($\theta \to \pi$). In this limit, the novel EEC observable in DIS is insensitive to pseudorapidity cuts, usually imposed in the forward and backward regions due to detector acceptance limitations. 

We obtained the singular distributions for EEC in DIS up to NNLO from the factorized formula and compared them against the full fixed-order QCD calculations up to NLO. The purpose of this comparison is twofold, first the numerical agreement in the $\theta \to \pi$ limit serves as a validation of our factorization formalism. Second, the point of deviation of the two distributions indicates the region where power corrections become relevant. Predictions up to N$^3$LL+NLO were presented.

Non-perturbative and hadronization effects for the EEC observable were investigated by considering  non-perturbative form factors extracted from the semi-inclusive hadron production in DIS.  Incorporating these non-perturbative models, we also presented the comparison of our predictions to \textsc{Pythia} simulations. 

Last but not least, we introduced a generalization of EEC in DIS where a subset of hadrons can be chosen. The modified observable can be used to tag the initial state and final state flavor, or for comparison against experimental measurements where only charged hadrons are considered. To conclude, we propose EEC in DIS as a way to check the universality of TMD factorization and study TMD PDFs and FFs with high precision, both experimentally and theoretically. We remark that EEC can be expanded for the study of spin effects in a polarized target hadron and therefore, constitutes a useful tool for the study TMD physics and nuclear matter effects in electron-ion collisions~\cite{Li:2020zbk,Li:2020rqj}  at the future EIC.\\

\begin{acknowledgements}
We would like to thank A. Bacchetta, R. Boughezal and F. Petriello for the comments on the manuscript. H.T.L. is supported by the DOE contract DE-AC02-06CH11357 and the NSF grant NSF-1740142. Y.M. is supported by the European Union’s Horizon 2020 research and innovation programme under the Marie Skłodowska-Curie grant agreement No. 754496 - FELLINI. I.V. is supported by the U.S. Department of Energy under Contract No. DE-AC52-06NA25396 and by the  LDRD program at LANL. \\
\end{acknowledgements}

\appendix
\section{Lorentz invariant definition of EEC in DIS}
\label{sec:BEEC-LI}
One can also write down a Lorentz invariant definition of the EEC in DIS. We denote this definition of the observable with EEC$_{\text{LI}}$ and it is given as follows,
\begin{equation}
     \text{EEC}_{\text{LI}} = \sum_a \int \frac{d\sigma_{\ell+ h \to \ell+ a+X}}{\sigma}\; z_a\; \delta(\tanh\bar{\eta}_a - \tanh\bar{\eta})\;,
\end{equation}
where
\begin{equation}
    \bar{\eta}_a = 2 \sqrt{1+ \frac{ q\cdot p_a}{x_B\, P \cdot p_a} } \;\;\;\xrightarrow[\text{frame}]{\text{Breit}} \;\;\; \frac{2 p_{a \perp}}{p_a^+}\;.
\end{equation}
For particles that satisfy the $\bar{n}$-collinear scaling as described in eq.~(\ref{eq:modes}) we have $\bar{\eta}_{\bar{n}} \simeq \pi -\theta_{\bar{n}}$. Therefore, in the back-to-back limit ($\bar{\eta} \to 0$), similarly to the case of \BEEC the new observable can be expressed in terms of the conventional TMD factorization.  
A advantage of using the above definition  is that, in contrast to the  one in eq.~(\ref{eq:definition-DIS}), it can be applied in the Laboratory frame and, thus, reduce the propagation of experimental uncertainties associated with the Lorentz transformation to the Breit frame.

\bibliography{main}
\bibliographystyle{JHEP}
\end{document}